\documentclass[12pt]{article}

\usepackage{cite}
\usepackage{epsfig}
\usepackage{amsmath}
\usepackage{array}
\usepackage{multirow}

\def\mathswitch#1{\relax\ifmmode#1\else$#1$\fi}
\def\mathswitchr#1{\relax\ifmmode{\mathrm{#1}}\else$\mathrm{#1}$\fi}
\newcommand{\PW}{\mathswitchr W}
\newcommand{\PZ}{\mathswitchr Z}

\newcommand{\Pt}{\mathswitchr t}

\newcommand{\MW}{\mathswitch {M_\PW}}
\newcommand{\MZ}{\mathswitch {M_\PZ}}

\newcommand{\mt}{\mathswitch {m_\Pt}}

\newcommand{\scrs}{\scriptscriptstyle}
\newcommand{\sw}{\mathswitch {s_{\scrs\PW}}}
\newcommand{\cw}{\mathswitch {c_{\scrs\PW}}}
\newcommand{\mw}{\mathswitch {\overline{M}_\PW}}
\newcommand{\mz}{\mathswitch {\overline{M}_\PZ}}

\newcommand{\gz}{\mathswitch {\overline{\Gamma}_\PZ}}

\newcommand{\seff}[1]{\sin^2\theta_{\rm eff}^{#1}}

\newcommand{\gev}{\,\, \mathrm{GeV}}
\newcommand{\mev}{\,\, \mathrm{MeV}}
\newcommand{\re}{\text{Re}\,}
\newcommand{\im}{\text{Im}\,}

\newcommand{\mycaption}[1]{\caption{\sl #1}}

\makeatletter
\def\section{\@startsection {section}{1}{\z@}{-3.5ex plus -1ex minus 
 -.2ex}{2.3ex plus .2ex}{\large\bf\boldmath}}
\def\subsection{\@startsection{subsection}{2}{\z@}{-3.25ex plus -1ex
 minus -.2ex}{1.5ex plus .2ex}{\normalsize\bf\boldmath}}
\def\subsubsection{\@startsection{subsubsection}{3}{\z@}{-3.25ex plus
 -1ex minus -.2ex}{1.5ex plus .2ex}{\normalsize\it}}

\graphicspath{{.}{plots/}}

\begin{document}
\thispagestyle{empty}

\def\thefootnote{\fnsymbol{footnote}}

\begin{flushright}
\end{flushright}

\vspace{1cm}

\begin{center}

{\Large {\bf Leading fermionic three-loop corrections to electroweak precision
observables}}
\\[3.5em]
{\large
Lisong Chen and Ayres~Freitas
}

\vspace*{1cm}

{\sl
Pittsburgh Particle-physics Astro-physics \& Cosmology Center
(PITT-PACC),\\ Department of Physics \& Astronomy, University of Pittsburgh,
Pittsburgh, PA 15260, USA
}

\end{center}

\vspace*{2.5cm}

\begin{abstract} Future electron-position colliders, such as the CEPC, FCC-ee, and ILC
have the capability to dramatically improve the experimental precision for W and
Z-boson masses and couplings. This would enable indirect probes of physics
beyond the Standard Model at multi-TeV scales. For this purpose, one must
complement the experimental measurements with equally precise calculations for
the theoretical predictions of these quantities within the Standard Model,
including three-loop electroweak corrections. This article reports on the
calculation of a subset of these corrections, stemming from diagrams with three
closed fermion loops to the following quantities: the prediction of the
W-boson mass from the Fermi constant, the effective weak mixing angle, and
partial and total widths of the Z boson. The numerical size of these corrections
is relatively modest, but non-negligible compared to the precision targets of
future colliders. In passing, an error is identified in previous results for
the two-loop corrections to the Z width, with a small yet non-zero numerical impact.
 \end{abstract}

\setcounter{page}{0}
\setcounter{footnote}{0}

\newpage


\section{Introduction}

Precision measurements of processes mediated by $W$ and $Z$ bosons are crucial
testbeds for the Standard Model (SM) and physics beyond the SM. Some of the most
important of these electroweak precision observables (EWPOs) are (a) muon decay,
mediated by a virtual $W$ boson, and (b) $e^+e^- \to f\bar{f}$, which is
primarily mediated by an $s$-channel $Z$ boson for center-of-mass energies
$\sqrt{s} \approx \MZ$. Here $f$ denotes any SM lepton or quark, except
the top quark. These processes receive sizable radiative corrections within the
SM, which are currently known at the full two-loop level
\cite{qcd2,mwshort,mwlong,mw,mwtot,swlept,swlept2,swbb,gz,zbos} and leading partial three- and four-loop results in powers of the top Yukawa coupling, $\alpha_t=\frac{y_t}{4\pi}$, which have been calculated at order $O(\alpha_t\alpha_s^2)$~\cite{qcd3} , $O(\alpha_t^2\alpha_s)$, $O(\alpha_t^3)$ ~\cite{mt6}, $O(\alpha_t\alpha_s^3)$ ~\cite{qcd4}. Including these corrections, the estimated theory
uncertainties from missing higher orders are safely below the current
experimental precision for these processes, see Refs.~\cite{rev1,rev2,pdg} for recent
reviews.

However, proposals for future high-luminosity $e^+e^-$ colliders, such as the CEPC
\cite{cepc}, FCC-ee \cite{fccee}, and ILC/Giga-Z~\cite{Irles:2019xny} would dramatically improve the experimental
precision for the relevant EWPOs, thus requiring significant additional
higher-order corrections to meet the physics goals \cite{therr}. In this article, we
report on the leading fermionic three-loop corrections to the EWPOs. Here
``leading fermionic'' refers to diagrams with the maximal number ($i.\,e.$
three) of closed fermion loops. Generally, contributions with closed fermion loops are numerically
enhanced since they are enhanced by powers of $\mt$ and a large number of
light fermion flavors. Technically, the leading fermionic corrections require
only the computation of one-loop integrals, but care has to be taken in the
derivation of the counterterms for the renormalization, as well as the
description of $e^+e^- \to f\bar{f}$ as a Laurent expansion about the complex
$Z$ pole \cite{zpole}.

%
%

%
Partial results for the leading fermionic three-loop corrections have been
discussed in Refs.~\cite{achim,georg98}, but the proper treatment of the complex gauge boson
pole was not addressed there.

In section~\ref{renorm}, the renormalization procedure and relevant counterterms are
discussed in more detail. Section~\ref{obs} describes the calculation of the leading
fermion three-loop corrections to the following quantities: (a) the Fermi
constant for muon decay, which can be used to predict the $W$ mass, 
(b) the effective weak mixing angle $\seff{f}$, which
describes the ratio of the vector and axial-vector couplings of the $Zf\bar{f}$
vertex, and (c) the partial widths for $Z \to f\bar{f}$. Numerical results are
presented in section~\ref{res}, together with a discussion of their impact.


\section{Renormalization}
\label{renorm}

The calculations presented in this article are based on the on-shell
renormalization scheme. In this scheme, the renormalized electromagnetic
coupling is defined through the electron-photon vertex at zero momentum
transfer, while the renormalized squared masses are defined at the real part of
the propagator poles. For particles with a non-negligible decay width, such as
the $W$ and $Z$ bosons, the propagator pole is complex and can be written as
\begin{align}
s_0 \equiv \overline{M}^2 - i\overline{M}\overline{\Gamma},
\end{align}
where $\overline{M}$ is the on-shell mass, while $\overline{\Gamma}$ is the
particle's decay width. This definition of the mass and width is rigorously
gauge-invariant \cite{zpole}, but it differs from the mass and width
commonly used in the literature. Denoting the latter by $M$ and $\Gamma$,
respectively, they are related according to
\begin{equation}
\textstyle
\overline{M} = M\big/\sqrt{1+\Gamma^2/M^2}\,, \qquad
\overline{\Gamma} = \Gamma\big/\sqrt{1+\Gamma^2/M^2}\,. \label{massrel}
\end{equation}
See $e.\,g.$ Refs.~\cite{mrel,rev1} for a more detailed discussion.

Including radiative corrections, the massive gauge boson two-point function
becomes
\begin{equation}
D(p^2) = p^2 - s_0 + \Sigma(p^2) - \delta \overline{M}^2\,,
\end{equation}
where $\Sigma(s)$ is the transverse part of the gauge boson self-energy,
and $\delta M^2$ is the mass counterterm. To avoid notational clutter, we do not
include a field or wavefunction renormalization for the gauge boson. Since
unstable particles can only appear as internal particles in a physical
process, any dependence on their field renormalization drops out in the
computation of such process\footnote{In our calculation, we have checked
explicitly that any field renormalization counterterms cancel.}.

In the on-shell scheme, $s_0$ is required to be a pole of the propagator,
$D(s_0)=0$. This leads to the conditions
\begin{align}
\delta\overline{M}^2 &= \re \Sigma\bigl(\overline{M}^2 -
i\overline{M}\overline{\Gamma}\bigr), \label{dmdef} \\
\overline{\Gamma} &= \frac{1}{\overline{M}} \, \im \Sigma\bigl(\overline{M}^2 -
i\overline{M}\overline{\Gamma}\bigr). \label{gamdef} 
\end{align}
By recursively inserting eq.~\eqref{gamdef} into \eqref{dmdef} and expanding in
orders of perturbation theory, the $W$-mass counterterm is given by
\begin{align}
\delta \overline{M}^2_{\PW(1)} = {}
 &\re \Sigma_{\PW(1)}(\mw^2)\,, \\[1ex]
\delta \overline{M}^2_{\PW(2)} = {}
 &\re \Sigma_{\PW(2)}(\mw^2) +
 \bigl[\im  \Sigma_{\PW(1)}(\mw^2)\bigr]
 \bigl[\im  \Sigma'_{\PW(1)}(\mw^2)\bigr]\,, \\[1ex]
\delta \overline{M}^2_{\PW(3)} = {}
 &\re \Sigma_{\PW(3)}(\mw^2) +
 \bigl[\im  \Sigma_{\PW(2)}(\mw^2)\bigr]
 \bigl[\im  \Sigma'_{\PW(1)}(\mw^2)\bigr] \notag \\
 &+\bigl[\im  \Sigma_{\PW(1)}(\mw^2)\bigr] \Bigl\{ \!
 \begin{aligned}[t]
  &\im  \Sigma'_{\PW(2)}(\mw^2)
  - \bigl[\im  \Sigma'_{\PW(1)}(\mw^2)\bigr]
   \bigl[\re  \Sigma'_{\PW(1)}(\mw^2)\bigr] \\
  &- \tfrac{1}{2}\bigl[\im  \Sigma_{\PW(1)}(\mw^2)\bigr]
   \bigl[\re  \Sigma''_{\PW(1)}(\mw^2)\bigr] \Bigr\}\,.
 \end{aligned}
\end{align}
Here and in the following the numbers in brackets denote the loop order.

For the $Z$-mass counterterm, one needs to include $\gamma$--$Z$ mixing effects.
The $Z$ and photon fields get renormalized according to
\begin{align}
Z_\mu &\to \sqrt{Z^{\PZ\PZ}}\,Z_\mu + \tfrac{1}{2}\delta Z^{\PZ\gamma}A_\mu\,,
\\
A_\mu &\to \tfrac{1}{2}\delta Z^{\gamma\PZ}Z_\mu +
 \sqrt{Z^{\gamma\gamma}}\,A_\mu \,.
\end{align}
As already mentioned above, in the following we will simply set
$Z^{\PZ\PZ},Z^{\gamma\gamma}\to 1$ for the $Z$ and photon field renormalization
constants, since these drop out anyways for the physical processes discussed in
this work. However, the mixing counterterms generate extra terms compared to
eqs.~\eqref{dmdef} and \eqref{gamdef}:
\begin{align}
\delta\mz^2 &= \re \Sigma_\PZ\bigl(\mz^2 -
i\mz\gz\bigr) + \tfrac{1}{4}\mz^2(\delta Z^{\gamma\PZ})^2, \label{dmzdef} \\
\gz &= \frac{1}{\mz[1+\frac{1}{4}(\delta Z^{\gamma\PZ})^2]} \, 
 \im \Sigma_\PZ\bigl(\mz^2 -i\mz\gz\bigr), \label{gzdef}
\intertext{where}
\Sigma_\PZ(p^2) &= \Sigma_{\PZ\PZ}(p^2) 
 - \frac{[\hat{\Sigma}_{\gamma\PZ}(p^2)]^2}{p^2+\hat{\Sigma}_{\gamma\gamma}(p^2)}
 \label{sigmaz}
 ,\\
\hat{\Sigma}_{\gamma\PZ}(p^2) &= \Sigma_{\gamma\PZ}(p^2) +
 \tfrac{1}{2}\delta Z^{\PZ\gamma}(p^2-\mz^2 - \delta\mz^2)
 + \tfrac{1}{2}\delta Z^{\gamma\PZ}p^2, \\
\hat{\Sigma}_{\gamma\gamma}(p^2) &= \Sigma_{\gamma\gamma}(p^2)
 + \tfrac{1}{4}(\delta Z^{\PZ\gamma})^2(p^2-\mz^2 - \delta\mz^2).
\end{align}
Here $\Sigma_{V_1V_2}$ is the self-energy with an incoming gauge boson $V_1$ and
outgoing gauge boson $V_2$.

\begin{figure}[tb]
\centering
\begin{tabular}{ll}
$\Sigma_{V_1V_2(1)} =$ &
\raisebox{-0.6cm}{\psfig{figure=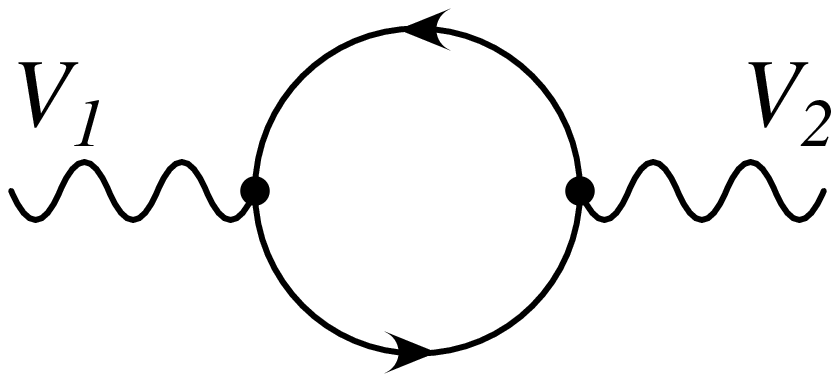, width=3.5cm}} \\[0.8cm]
$\Sigma_{V_1V_2(2)} =$ &
\raisebox{-0.6cm}{\psfig{figure=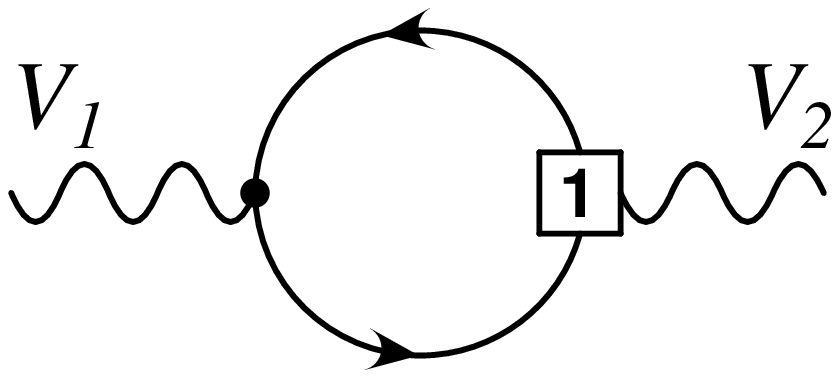, width=3.5cm}} +
\raisebox{-0.6cm}{\psfig{figure=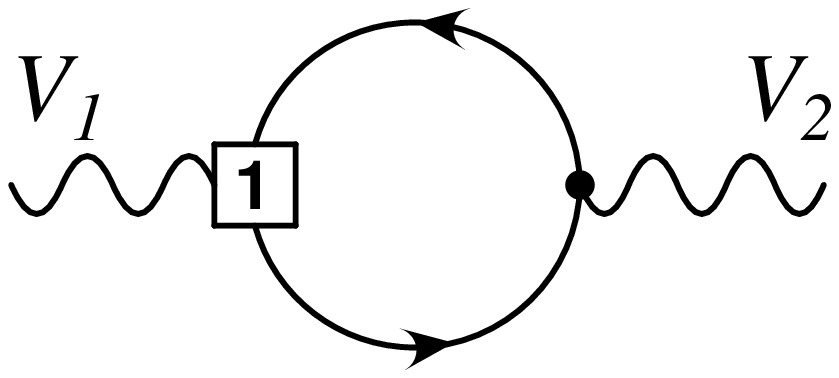, width=3.5cm}} \\[0.8cm]
$\Sigma_{V_1V_2(3)} =$ &
\raisebox{-0.6cm}{\psfig{figure=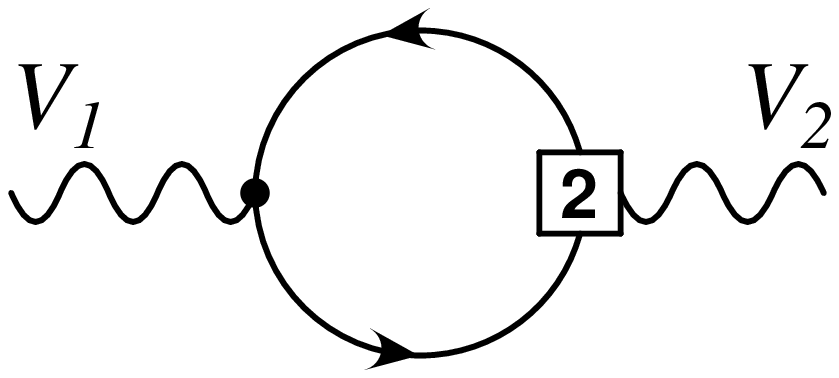, width=3.5cm}} +
\raisebox{-0.6cm}{\psfig{figure=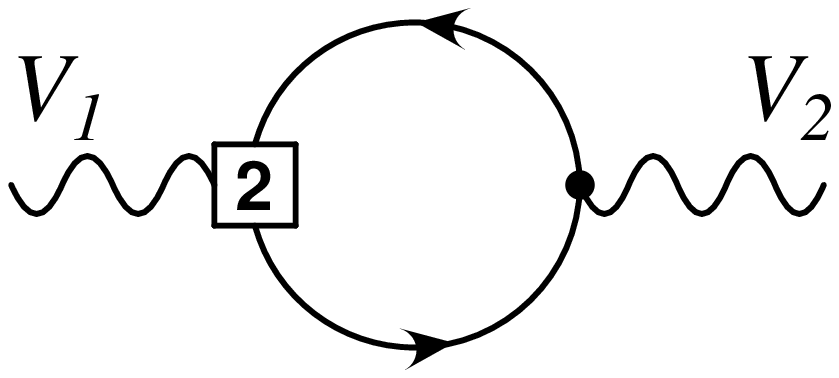, width=3.5cm}} +
\raisebox{-0.6cm}{\psfig{figure=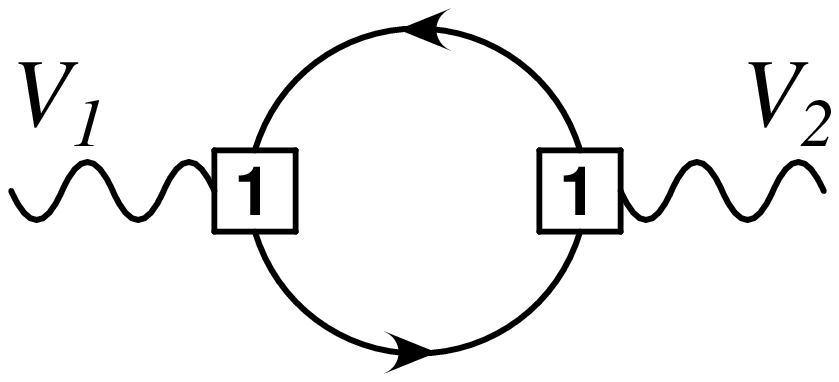, width=3.5cm}} 
\end{tabular}
\vspace{1ex}
\mycaption{Diagrams with closed fermion loops contributing to self-energies at
different orders. A box with number $n$ indicates a counterterm of loop
order $n$.One should notice that there are no one-particle irreducible diagrams with two or three explicit closed fermion loops.
\label{fig:sesub}}
\end{figure}

\begin{figure}[tb]
\centering
\begin{tabular}{l@{}l}
\raisebox{-0.9cm}{\psfig{figure=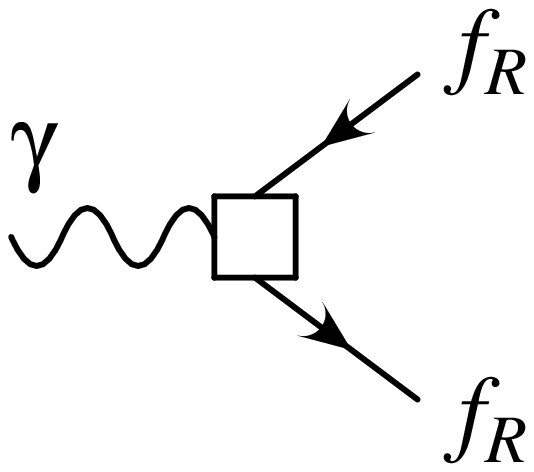, width=2.3cm}} &
$= -i\gamma_\mu \, e(1+\delta Z_e) \, Q_f \bigl(1 + \frac{\sw+\delta\sw}{2(\cw+\delta\cw)}
 \delta Z^{\PZ\gamma}\bigr)$ \\
\raisebox{-0.9cm}{\psfig{figure=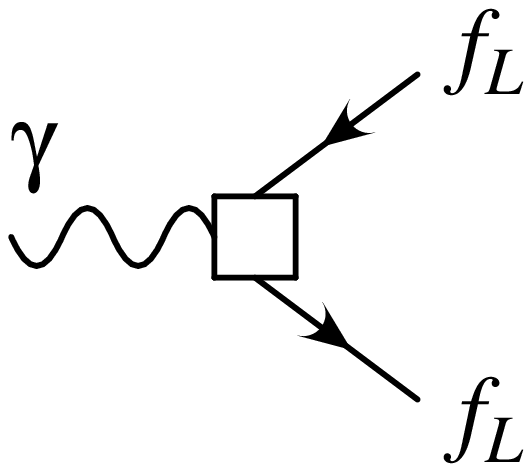, width=2.3cm}} &
$= -i\gamma_\mu \, e(1+\delta Z_e) \, \Bigl[ Q_f \bigl(1 + \frac{\sw+\delta\sw}{2(\cw+\delta\cw)}
 \delta Z^{\PZ\gamma}\bigr) - \frac{I^3_f}{2(\sw+\delta\sw)(\cw+\delta\cw)}\delta Z^{\PZ\gamma}
 \Bigr]$ \\
\raisebox{-0.9cm}{\psfig{figure=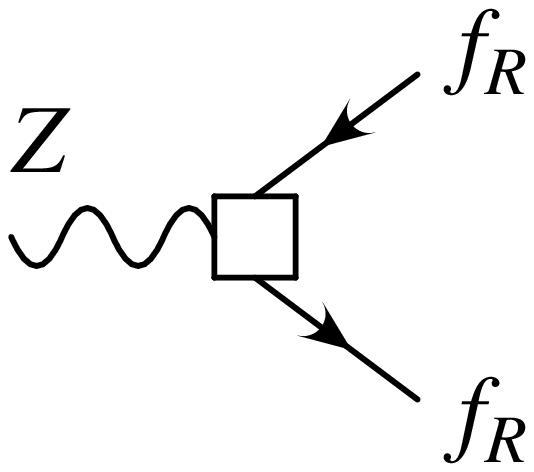, width=2.3cm}} &
$= -i\gamma_\mu \, e(1+\delta Z_e) \, Q_f \bigl(\frac{\sw+\delta\sw}{(\cw+\delta\cw)}
 + \frac{1}{2}\delta Z^{\gamma\PZ} \bigr)$ \\
\raisebox{-0.9cm}{\psfig{figure=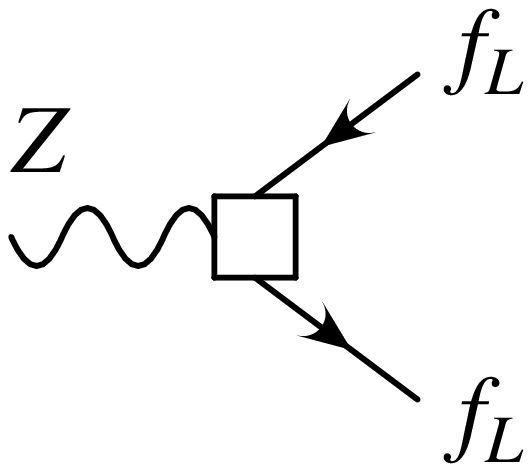, width=2.3cm}} &
$= -i\gamma_\mu \, e(1+\delta Z_e) \, \Bigl[ Q_f \bigl(\frac{\sw+\delta\sw}{(\cw+\delta\cw)}
 + \frac{1}{2}\delta Z^{\gamma\PZ} \bigr) - \frac{I^3_f}{(\sw+\delta\sw)(\cw+\delta\cw)}
 \Bigr]$ \\
\raisebox{-0.9cm}{\psfig{figure=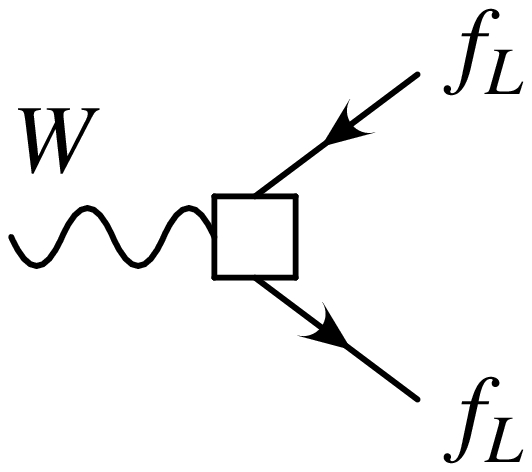, width=2.3cm}} &
$= i\gamma_\mu \, \frac{e(1+\delta Z_e)}{\sqrt{2}(\sw+\delta\sw)}$
\end{tabular}
\mycaption{Feynman rules with counterterms for gauge-boson--fermion vertices.
Here $I^3_f = \pm \frac{1}{2}$ for up/down-type fermions, and $Q_f$ is the
fermion electric charge in units of $e$.
$\delta Z_e$ and $\delta\sw$ are the change and weak mixing angle
counterterms, respectively. Furthermore, $\cw+\delta\cw =
\sqrt{1-(\sw+\delta\sw)^2}$.
Note that field or wavefunction renormalization counterterms have not been 
included, since they are either irrelevant for the processes considered here, as 
explained in the text, or do not receive any contributions from closed fermion
loops.
\label{fig:rules}}
\end{figure}

The mixing counterterms are fixed through the conditions
\begin{equation}
\hat{\Sigma}_{\gamma\PZ}(0)=0, \qquad
\re \hat{\Sigma}_{\gamma\PZ}\bigl(\mz^2-i\mz\gz\bigr) = 0. \label{mixcond}
\end{equation}
Using eqs.~\eqref{dmzdef}--\eqref{mixcond} and expanding in orders of
perturbation theory yields
\begin{align}
\delta \overline{M}^2_{\PZ(1)} = {}
 &\re \Sigma_{\PZ\PZ(1)}(\mz^2) \displaybreak[0] \\[1ex]
\delta \overline{M}^2_{\PZ(2)} = {}
 &\re \Sigma_{\PZ\PZ(2)}(\mz^2)  +
 \bigl[\im  \Sigma_{\PZ\PZ(1)}(\mz^2)\bigr]
 \bigl[\im  \Sigma'_{\PZ\PZ(1)}(\mz^2)\bigr] \notag \\
 & + \frac{\bigl[\im  \Sigma_{\gamma\PZ(1)}(\mz^2)\bigr]^2}{\MZ^2}
 + \tfrac{1}{4}\mz^2 \,(\delta Z^{\gamma\PZ}_{(1)})^2 \displaybreak[0] \\[1ex]
\delta \overline{M}^2_{\PZ(3)} = {}
 &\re \Sigma_{\PZ\PZ(3)}(\mz^2) +
 \bigl[\im  \Sigma_{\PZ\PZ(2)}(\mz^2)\bigr]
 \bigl[\im  \Sigma'_{\PZ\PZ(1)}(\mz^2)\bigr] \notag \\
 &+\bigl[\im  \Sigma_{\PZ\PZ(1)}(\mz^2)\bigr] \Bigl\{ \!
 \begin{aligned}[t]
  &\im  \Sigma'_{\PZ\PZ(2)}(\mz^2)
  - \bigl[\im  \Sigma'_{\PZ\PZ(1)}(\mz^2)\bigr]
   \bigl[\re  \Sigma'_{\PZ\PZ(1)}(\mz^2)\bigr] \\
  &- \tfrac{1}{2}\bigl[\im  \Sigma_{\PZ\PZ(1)}(\mz^2)\bigr]
   \bigl[\re  \Sigma''_{\PZ\PZ(1)}(\mz^2)\bigr] \\
  &- \frac{\im  \Sigma_{\gamma\PZ(1)}(\mz^2)}{\mz^2}
   \bigl[2\,\re  \Sigma'_{\gamma\PZ(1)}(\mz^2) + \delta Z^{\gamma\PZ}_{(1)}
   + \delta Z^{\PZ\gamma}_{(1)}\bigr] \Bigr\}  
 \end{aligned} \notag \\[-1ex]
 &+\frac{\im  \Sigma_{\gamma\PZ(1)}(\mz^2)}{\mz^2} \Bigl\{ \!
 \begin{aligned}[t]
   2\,\im  \Sigma_{\gamma\PZ(2)}(\mz^2) - 
   \frac{\im  \Sigma_{\gamma\PZ(1)}(\mz^2)}{\mz^2}
   \bigl[\im  \Sigma_{\gamma\gamma(1)}(\mz^2)\bigr] \Bigr\}
 \end{aligned} \notag \\
 &+ \tfrac{1}{2}\mz^2 \,\delta Z^{\gamma\PZ}_{(1)}\,\delta Z^{\gamma\PZ}_{(2)}
 \label{dmz3} \,.
\end{align}
The self-energies receive contributions from one-loop diagrams with counterterm
insertions, see Fig.~\ref{fig:sesub},
with relevant counterterm Feynman rules shown in Fig.~\ref{fig:rules}.
It is worth noting that when inserting these into eq.~\eqref{dmz3}, $\delta
Z^{\gamma\PZ}_{(2)}$ drops out without needing to include
an explicit expression for it.


Other relevant counterterms are given by
\begin{align}
\delta Z^{\PZ\gamma}_{(n)} &= 0, \label{dzza} \\
\delta Z_{e(1)} &= \frac{\alpha}{9\pi}\biggl[\frac{12}{\epsilon} + \frac{50}{3} 
 - 2\,L(\mt^2)-10\,L(\MZ^2) \biggr] + \frac{\Delta\alpha}{2}\,, \label{dze1} \\ 
\delta Z_{e(2)} &= \frac{3}{2}(\delta Z_{e(1)})^2, \\
\delta Z_{e(3)} &= \frac{5}{2}(\delta Z_{e(1)})^3, \label{dze3} \\
\sw+\delta\sw &= \sqrt{1-\frac{\mw^2+\delta\mw^2}{\mz^2+\delta\mz^2}} \label{dsw}
\end{align}
with $L(m^2) \equiv \ln \frac{m^2}{4\pi\mu^2}+\gamma_{\rm E}$.
The simple results in eqs.~\eqref{dzza}--\eqref{dze3} are a consequence of restricting
ourselves to only closed fermion loop. The weak mixing angle counterterm is
obtained by demanding that the relation $\sw^2 = 1-\mw^2/\mz^2$ holds to all
orders. Order-by-order expressions for
$\delta\sw{}_{(n)}$ can be obtained by plugging the previous expressions for the
mass counterterms into \eqref{dsw}, but we refrain from spelling them out here.

The symbol $\Delta\alpha$ in eq.~\eqref{dze1} stems from light-fermion loop
contributions in the photon vacuum polarization,
\begin{equation}
\Delta\alpha =
\Pi^{\gamma\gamma}_{\rm lf}(\MZ^2)-\Pi^{\gamma\gamma}_{\rm lf}(0), \qquad 
\text{where} \quad
\Pi^{\gamma\gamma}(q^2) = \frac{\Sigma^{\gamma\gamma}(q^2)}{q^2}.
\label{delalph}
\end{equation}
$\Pi^{\gamma\gamma}_{\rm lf}(q^2)$ can be divided into a leptonic part, which is
perturbatively calculable \cite{dalept}, and a hadronic part that becomes non-perturbative for
small $q^2$. Therefore the hadronic contribution is commonly extracted from data
\cite{dahad}. When using results from the literature for $\Delta\alpha_{\rm
had}$, one must remember that these references use $\MZ$ rather $\mz$ in
eq.~\eqref{delalph}. Accordingly, one must also use $\MZ$ rather $\mz$ in
eq.~\eqref{dze1}.


\section{Definition of the observables}
\label{obs}

\subsection{Fermi constant \boldmath $G_\mu$}

The Fermi constant can be determined with high precision from the muon decay
lifetime \cite{Tishchenko:2012ie}. In the SM it is defined through
\begin{equation}
G_\mu = \frac{\pi\alpha}{\sqrt{2}\sw^2\mw^2}(1+\Delta r), \label{gmu}
\end{equation}
where $\Delta r$ includes the contribution from radiative corrections.
With $\sw^2=1-\mw^2/\mz^2$, eq.~\eqref{gmu} can be used to compute a prediction
for the $W$-boson mass in terms of $G_\mu$ and other SM parameters. Since
$\Delta r$ itself depends on $\mw$ and $\sw$, this is usually performed in a
recursive procedure.

The corrections to $\Delta r$ from diagrams with closed fermion loops can be
written as
\begin{equation}
1+\Delta r = \biggl( \frac{1+\delta Z_e}{\sw+\delta\sw} \biggr)^2
 \frac{\mw^2}{\mw^2+\delta\mw^2 - \Sigma_\PW(0)}\,. \label{dr}
\end{equation}
It is permissible to set the momentum transfer in the $W$ propagator and in
$\Sigma_\PW$ to zero since $m_\mu \ll \mw$.
Eq.~\eqref{dr} can be computed straightforwardly by expanding in orders of perturbation
theory and using the derivations from section~\ref{renorm}.

\subsection{Effective weak mixing angle \boldmath $\seff{f}$}

\begin{figure}[tb]
\centering
\begin{tabular}{ll}
\raisebox{-1.1cm}{\psfig{figure=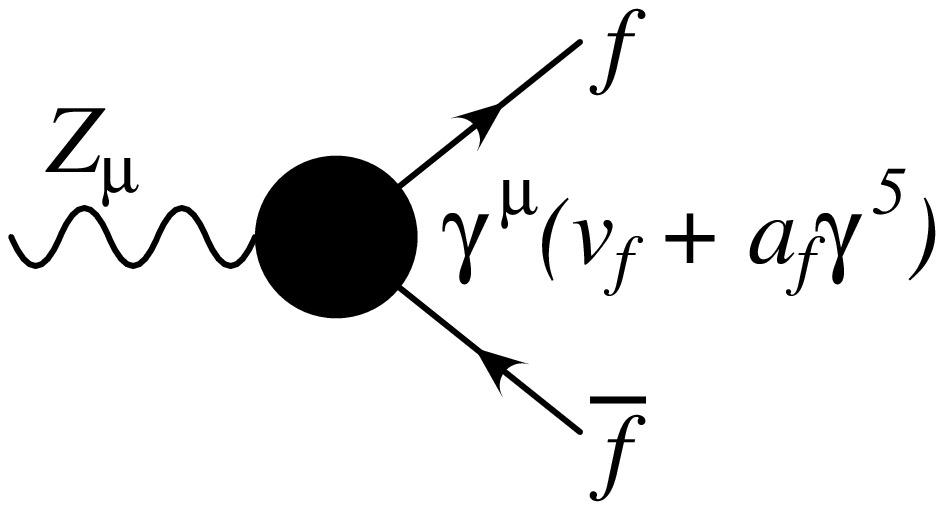, height=2.2cm}} = &
 \raisebox{-0.75cm}{\psfig{figure=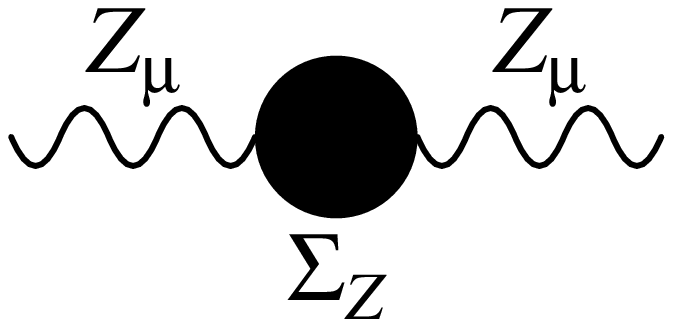, height=1.5cm}} = \\
\phantom{+} \psfig{figure=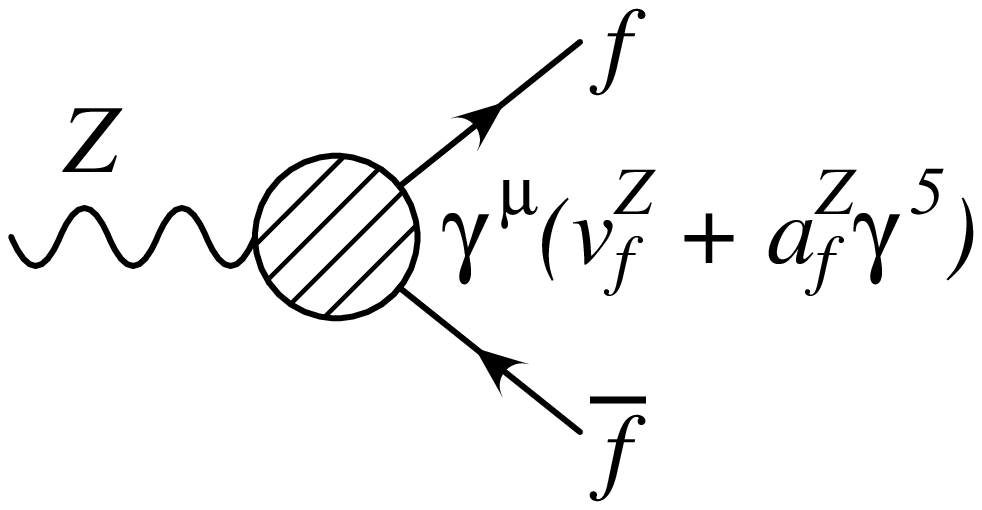, height=2.2cm} &
 \phantom{+} \raisebox{0.35cm}{\psfig{figure=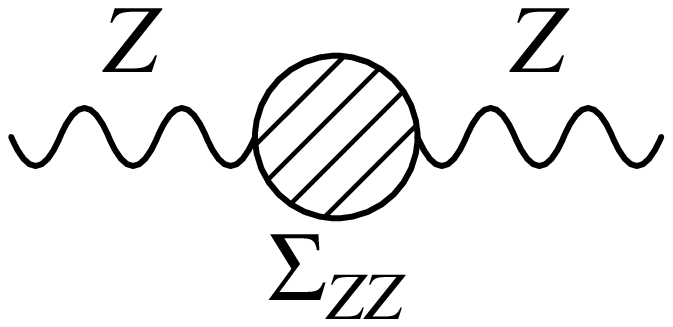, height=1.5cm}} \\
+ \raisebox{-1.1cm}{\psfig{figure=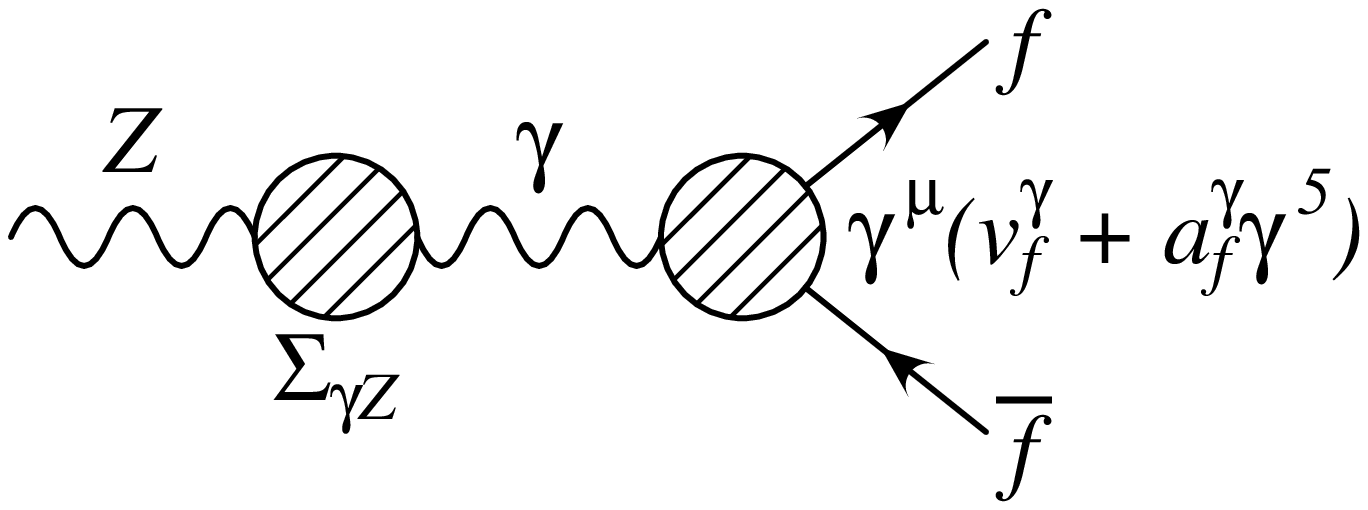, height=2.2cm}} &
 + \raisebox{-0.75cm}{\psfig{figure=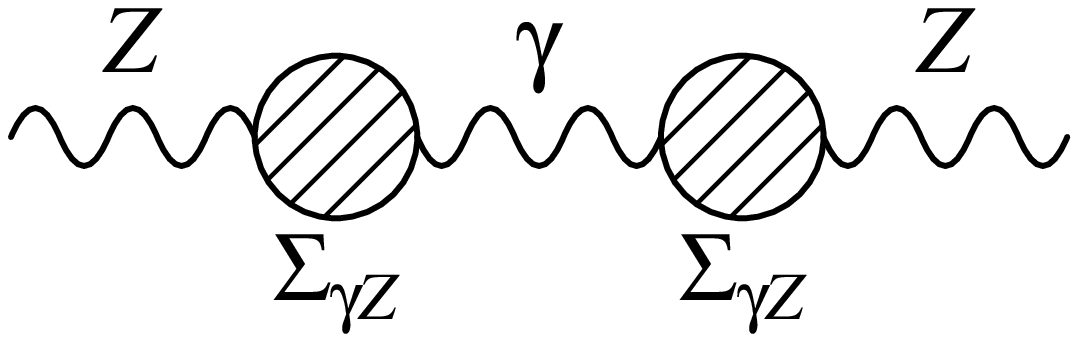, height=1.5cm}} \\ 
+ \raisebox{-1.1cm}{\psfig{figure=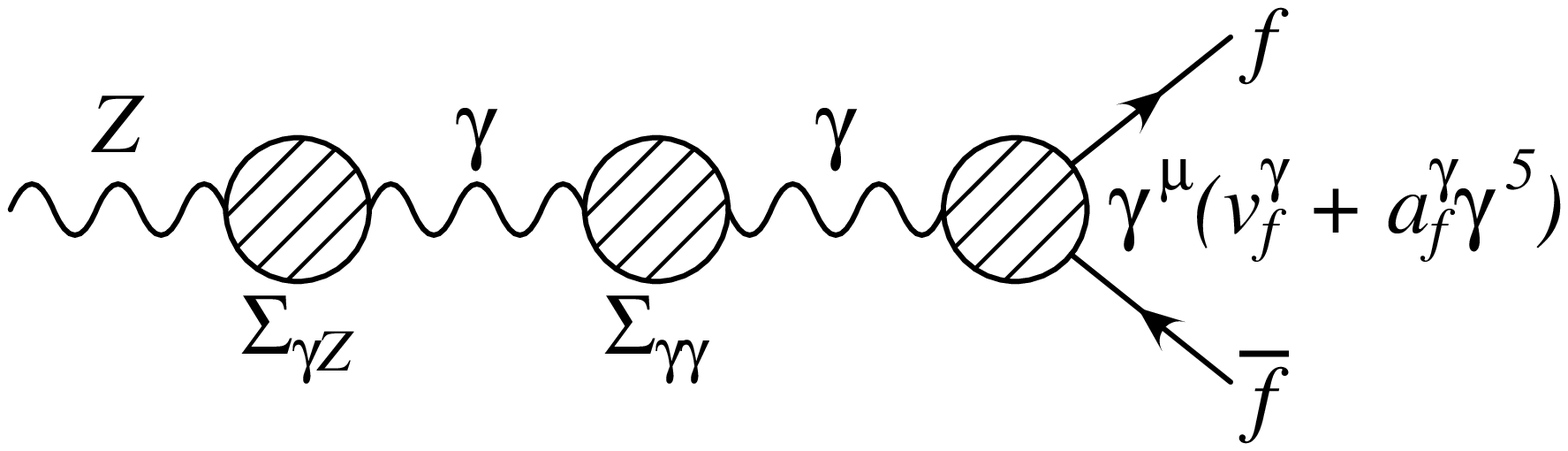, height=2.2cm}}\hspace{-2cm} &
 + \raisebox{-0.75cm}{\psfig{figure=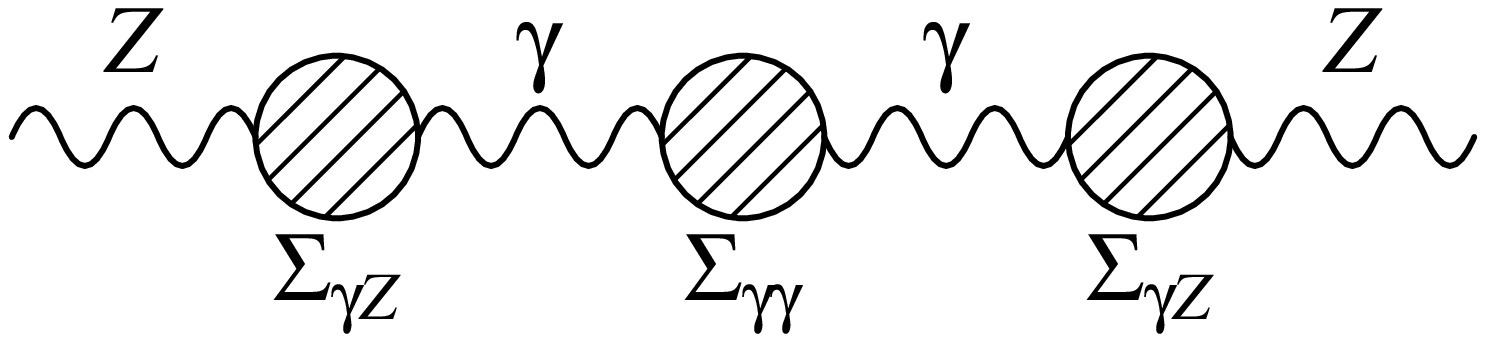, height=1.5cm}} \\ 
+ \dots & + \dots
\end{tabular}
\mycaption{Decomposition of the effective $Zf\bar{f}$ vertex and $Z$ self-energy
into one-particle irreducible building blocks, indicated by the
hatched blobs.
\label{fig:azmix}}
\end{figure}

The effective weak mixing angle is defined in terms of the effective vector and
axial-vector couplings of the $Z$-boson to an $f\bar{f}$ pair, denoted $v_f$ and
$a_f$, respectively,
\begin{equation}
\seff{f} = \frac{1}{4|Q_f|}\Bigl(1+\re\frac{v_f}{a_f}\Bigr)_{s=\mz^2}\,.
\end{equation}
When considering corrections with closed fermion loops, the effective couplings
are obtained from the relations
\begin{align}
a_f(s) &= -\frac{e(1+\delta Z_e)I_3^f}{2(\sw+\delta\sw)(\cw+\delta\cw)}\,, 
 \label{aeff} \\
v_f(s) &= \frac{e(1+\delta Z_e)[I_3^f - 2Q_f(\sw+\delta\sw)^2]}{%
	    2(\sw+\delta\sw)(\cw+\delta\cw)}
	  + e(1+\delta Z_e)Q_f \frac{\Sigma_{\gamma\PZ}(s) 
	   - \frac{1}{2}\delta Z^{\gamma Z}\Sigma_{\gamma\gamma}(s)}{%
	   s+\Sigma_{\gamma\gamma}(s)}\,, \label{veff}
\intertext{where}
&\cw+\delta\cw = \sqrt{1-(\sw+\delta\sw)^2}
\end{align}
In writing these equations, we have used that $\delta Z^{\PZ\gamma}=0$, see
eq.~\eqref{dzza}, and set $Z^{\PZ\PZ},Z^{\gamma\gamma}\to 1$ as discussed above.
The second term in \eqref{veff} originates from photon-$Z$ mixing self-energies,
as illustrated in Fig.~\ref{fig:azmix}.

\subsection{Partial width \boldmath $\Gamma[Z\to f\bar{f}]$}

Recursively expanding eq.~\eqref{gzdef} in loop orders yields
\begin{align}
\gz = \frac{1}{\mz}\,
\Bigl \{&\im \Sigma_{\PZ(1)} 
\;+\; \im \Sigma_{\PZ(2)} - 
 (\im \Sigma_{\PZ(1)})(\re \Sigma'_{\PZ(1)}) \notag\\
&+ \im \Sigma_{\PZ(3)} - (\im \Sigma_{\PZ(2)})(\re \Sigma'_{\PZ(1)}) \notag\\
&\quad+ (\im \Sigma_{\PZ(1)})
  \bigl [(\re \Sigma'_{\PZ(1)})^2-\re \Sigma'_{\PZ(2)} - \tfrac{1}{4}
  (\delta Z^{\gamma\PZ}_{(1)})^2 
  -\tfrac{1}{2}(\im \Sigma_{\PZ(1)})(\im \Sigma''_{\PZ(1)}) \bigr] \notag\\[1ex]
&+ \im \Sigma_{\PZ(4)} - (\im \Sigma_{\PZ(3)})(\re \Sigma'_{\PZ(1)}) \notag\\
&\quad+ (\im \Sigma_{\PZ(2)})
  \bigl [(\re \Sigma'_{\PZ(1)})^2-\re \Sigma'_{\PZ(2)} - \tfrac{1}{4}
  (\delta Z^{\gamma\PZ}_{(1)})^2 
  -(\im \Sigma_{\PZ(1)})(\im \Sigma''_{\PZ(1)}) \bigr] \notag\\
&\quad+ (\im \Sigma_{\PZ(1)})
  \bigl [-(\re \Sigma'_{\PZ(1)})^3
   +2(\re \Sigma'_{\PZ(2)})(\re \Sigma'_{\PZ(1)})
   - \re \Sigma'_{\PZ(3)} \notag\\
&\qquad- \tfrac{1}{2} \delta Z^{\gamma\PZ}_{(1)}\,\delta Z^{\gamma\PZ}_{(2)}
 +\tfrac{1}{2}(\re \Sigma'_{\PZ(1)})(\delta Z^{\gamma\PZ}_{(1)})^2
  -\tfrac{1}{2}(\im \Sigma_{\PZ(1)})(\im \Sigma''_{\PZ(2)}) \notag\\
&\qquad+ \tfrac{3}{2}(\im \Sigma_{\PZ(1)})(\re \Sigma'_{\PZ(1)})(\im \Sigma''_{\PZ(1)})
 + \tfrac{1}{6}(\im \Sigma_{\PZ(1)})^2(\re \Sigma'''_{\PZ(1)})\bigr] 
\Bigr \}_{s=\mz^2}\;.  
\end{align}
When neglecting light fermion masses, $\im \Sigma''_\PZ = 0$. Note that
$\Sigma_\PZ$ includes $Z$--$\gamma$ mixing effects, see eq.~\eqref{sigmaz}. One can easily extend such an expansion up to higher orders.

Through the optical theorem, the imaginary part of $\Sigma_\PZ$ can be expressed
in terms of the decay rate for $Z \to f\bar{f}$,
\begin{equation}
\im \Sigma_{\PZ} = \frac{1}{3\mz} \sum_{f} \sum_{\rm
spins} \int d\Phi \;\bigl (|v_f|^2 + |a_f|^2\bigr ),
\end{equation}
where $a_f$ and $v_f$ are the effective axial-vector and vector couplings,
defined in eqs.~\eqref{aeff} and \eqref{veff}, respectively.
Then one obtains the following expressions for the total and partial widths of
the Z boson:
\begin{align}
\gz &= \sum_f \overline{\Gamma}_f\,, \qquad
\overline{\Gamma}_f = \frac{N_c^f\mz}{12\pi} \Bigl [
 {\cal R}_{\rm V}^f F_{\rm V}^f + {\cal R}_{\rm A}^f F_{\rm A}^f \Bigr ]_{s=\mz^2} 
 \;, \label{Gz} \displaybreak[0] \\
F_{\rm V}^f &= v_{f(0)}^2  + 2 \,\re  (v_{f(0)}v_{f(1)}) - v_{f(0)}^2\,\re \Sigma'_{\PZ(1)} 
 \notag \\[.5ex]
&\quad + 2 \,\re  (v_{f(0)}v_{f(2)}) + |v_{f(1)}|^2 
 - 2 \,\re  (v_{f(0)}v_{f(1)})\,\re \Sigma'_{\PZ(1)} \notag \\
&\qquad + v_{f(0)}^2 \bigl [(\re \Sigma'_{\PZ(1)})^2-\re \Sigma'_{\PZ(2)}
 - \tfrac{1}{4} (\delta Z^{\gamma\PZ}_{(1)})^2
 - \tfrac{1}{2}(\im \Sigma_{\PZ(1)})(\im \Sigma''_{\PZ(1)}) \bigr ] \notag
 \displaybreak[0] \\[.5ex]
&\quad + 2 \,\re  (v_{f(0)}v_{f(3)} + v_{f(1)}^*v_{f(2)})
 - \bigl[ 2 \,\re  (v_{f(0)}v_{f(2)}) + |v_{f(1)}|^2 \bigr] 
 \,\re \Sigma'_{\PZ(1)} \notag \\
&\qquad +2 \,\re  (v_{f(0)}v_{f(1)})\, \bigl [(\re \Sigma'_{\PZ(1)})^2-\re \Sigma'_{\PZ(2)}
 - \tfrac{1}{4} (\delta Z^{\gamma\PZ}_{(1)})^2
 - (\im \Sigma_{\PZ(1)})(\im \Sigma''_{\PZ(1)}) \bigr ] \notag \\
&\qquad+ v_{f(0)}^2
  \bigl [-(\re \Sigma'_{\PZ(1)})^3
   +2(\re \Sigma'_{\PZ(2)})(\re \Sigma'_{\PZ(1)})
   - \re \Sigma'_{\PZ(3)} \notag\\
&\qquad\qquad\quad- \tfrac{1}{2} \delta Z^{\gamma\PZ}_{(1)}\,\delta Z^{\gamma\PZ}_{(2)}
 +\tfrac{1}{2}(\re \Sigma'_{\PZ(1)})(\delta Z^{\gamma\PZ}_{(1)})^2
  -\tfrac{1}{2}(\im \Sigma_{\PZ(1)})(\im \Sigma''_{\PZ(2)}) \notag\\
&\qquad\qquad\quad+ \tfrac{3}{2}(\im \Sigma_{\PZ(1)})(\re \Sigma'_{\PZ(1)})(\im \Sigma''_{\PZ(1)})
 + \tfrac{1}{6}(\im \Sigma_{\PZ(1)})^2(\re \Sigma'''_{\PZ(1)})\bigr] 
 , \label{Fv} 
\end{align}
and analogously for $F_{\rm A}^f$. Here $N_c^f = 3(1)$ for quarks (leptons), and 
the functions ${\cal R}_{\rm V,A}$ are included in general to account for
final-state QCD and QED corrections. When considering corrections with only
closed fermion loops, ${\cal R}_{\rm V,A}=1$.

\subsection{Technical aspects of the calculation} 
\label{calc}

Since the algebraic expressions for our results are rather lengthy, the
calculation has been carried out with the help of computer algrebra tools,
within the framework of {\sc Mathematica}. {\sc FeynArts 3.3} \cite{feynarts}
has been used for the generation of diagrams and amplitudes, and {\sc FeynCalc
9.2.0} \cite{feyncalc} has been employed for some of the Dirac and tensor algebra.
The masses and Yukawa couplings of all fermions except the top quark have been
neglected. Furthermore, CKM mixing of the top quark with other quark generations
has been ignored.

\bigskip
We have compared results for $\Delta r$, $\seff{f}$ and $\overline{\Gamma}_f$
with two fermionic loops with Refs.~\cite{mwshort,mwlong}, \cite{swlept} and
\cite{gz}, respectively. Exact algebraic agreement was found, with one
exception: The $\re \Sigma'_{\PZ(2)}$ term in the third line of \eqref{Fv},
together with \eqref{sigmaz} and the $\gamma$--$Z$ mixing counterterms, leads to
\begin{equation}
\re \Sigma'_{\PZ\PZ(2)}(s) - \frac{d}{ds}\, \biggl(
 \frac{[\im\Sigma_{\gamma\PZ(1)}(s)]^2}{s}\biggr) \label{rezp2}
\end{equation}
The second term in eq.~\eqref{rezp2} was missed in Ref.~\cite{gz}. The numerical
impact of this term will be discussed in the following section.


\section{Numerical results}
\label{res}

Let us now discuss the numerical impact of the leading fermionic three-loop
corrections to the observables introduced in the previous section. For
concreteness, input parameters in Tab.~\ref{tab:input} are used, but the results
do not depend very strongly on the specific input values within experimentally
allowed ranges.

\begin{table}[tb]
\renewcommand{\arraystretch}{1.2}
\begin{center}
\begin{tabular}{|r@{$\;=\;$}ll|}
\hline
$\MZ$ & $91.1876\gev$ & \multirow{2}{*}{$\biggr\}\!\Rightarrow\;
 \mz = 91.1535\gev$} \\
$\Gamma_\PZ$ & $2.4952\gev$ & \\
$\MW$ & $80.358\gev$ & \multirow{2}{*}{$\biggr\}\!\Rightarrow\;
 \mw = 80.331\gev$} \\
$\Gamma_\PW$ & $2.089\gev$ & \\
$\mt$ & $173.0\gev$ & \\
$m_{f\neq \Pt}$ & 0 & \\
$\alpha$ & \multicolumn{2}{@{}l|}{$1/137.035999084$} \\
$\Delta\alpha$ & $0.05900$ & \\
$G_\mu$ & \multicolumn{2}{@{}l|}{$1.1663787 \times 10^{-5}\gev^{-2}$} \\
\hline
\end{tabular}
\end{center}
\vspace{-2ex}
\mycaption{Benchmark values for the input parameters used in the numerical
analysis, based on Ref.~\cite{pdg}.
\label{tab:input}}
\end{table}

With these inputs, the fermionic three-loop corrections to $\Delta r$ is found
to be
\begin{align}
\Delta r_{(3)} = 2.50 \times 10^{-5}.
\end{align}
This can be translated into a shift, $\Delta M_{\PW(3)}$, of the predicted value
of the $W$-boson mass in the SM, using eq.~\eqref{gmu}. Given that $\Delta
r_{(3)}$, it is sufficient to expand eq.~\eqref{gmu} up to linear order in
$\Delta r_{(3)}$ and $\Delta M_{\PW(3)}$, leading to
\begin{align}
\Delta \overline{M}_{\PW(3)} \approx \frac{\pi \alpha \mz^2}{2\sqrt{2}G_\mu \mw
 (\mz^2-2\mw^2)} \,\Delta r_{(3)}
 = -0.389\mev. \label{dmw}
 \end{align}

For the effective weak mixing angle, the fermionic three-loop correction amounts
to
\begin{align}
\Delta\sin^2\theta^f_{\rm eff,(3)} &= 1.34 \times 10^{-5}
&& [\MW \text{ as indep.\ input}].
\intertext{%
This result does not depend on the type of fermion $f$. If we assume that $\MW$
is predicted from $G_\mu$, we can take into account the leading effect of
the shift $\Delta M_{\PW(3)}$ from eq.~\eqref{gmu} according to}
\Delta'\sin^2\theta^f_{\rm eff,(3)} &= 
 \Delta\sin^2\theta^f_{\rm eff,(3)} - \frac{\Delta \overline{M}_{\PW(3)}^2}{\mz^2}
 = 2.09 \times 10^{-5} \label{dseff}
&& [\MW \text{ from } G_\mu].
\end{align}
In a similar fashion, one obtains the corrections to the partial decay widths
\begin{align}
&\begin{aligned}[c]
&\Delta\overline{\Gamma}_{f,(3)} = N_c^f \bigl[0.105\, (I_3^f)^2
 - 0.105 \, I_3^f Q_f + 0.046 \, Q_f^2 \bigr] \mev, \\
&\Delta\overline{\Gamma}_{\ell,(3)} = 0.019\mev, \\
&\Delta\overline{\Gamma}_{\nu,(3)} = 0.026\mev, \\
&\Delta\overline{\Gamma}_{\rm d,(3)} = 0.041\mev, \\
&\Delta\overline{\Gamma}_{\rm u,(3)} = 0.035\mev, \\
&\Delta\overline{\Gamma}_{\rm tot,(3)} = 0.331\mev,
\end{aligned}
&& [\MW \text{ as indep.\ input}] \label{dgamfx} \displaybreak[0] \\[1ex]
&\Delta'\overline{\Gamma}_{f,(3)} = 
 \Delta\overline{\Gamma}_{f,(3)} - \frac{\Delta\overline{M}_{\PW(3)}^2}{\mz}
 \times \frac{\alpha N_c^f}{6\sw^4\cw^4} \bigl[(2\sw^2-1)(I_3^f)^2
  + 2\sw^4Q_f(Q_f-I_3^f)\bigr] \hspace{-8em} \displaybreak[0] \\[1ex]
&\begin{aligned}[c]
&\Delta'\overline{\Gamma}_{f,(3)} = N_c^f \bigl[0.090\, (I_3^f)^2
 - 0.108 \, I_3^f Q_f + 0.048 \, Q_f^2 \bigr] \mev, \\
&\Delta'\overline{\Gamma}_{\ell,(3)} = 0.017\mev, \\
&\Delta'\overline{\Gamma}_{\nu,(3)} = 0.022\mev, \\
&\Delta'\overline{\Gamma}_{\rm d,(3)} = 0.029\mev, \\
&\Delta'\overline{\Gamma}_{\rm u,(3)} = 0.024\mev, \\
&\Delta'\overline{\Gamma}_{\rm tot,(3)} = 0.255\mev.
\end{aligned}
&& [\MW \text{ from } G_\mu] \label{dgamf} 
\end{align}
The results in eqs.~\eqref{dmw}, \eqref{dgamfx} and \eqref{dgamf} are presented
in terms of the gauge-invariant complex-pole definitions of the gauge-boson
masses and widths. However, the corresponding corrections to the
conventional (unbarred) masses and widths are the same within the precision
presented above, since the translation factor in eq.~\ref{massrel} is about
1.00035, $i.\,e.$ very close to 1.

When comparing the above results with the experimental determination of these
quantities \cite{pdg}, which are dominated by measurements from LEP, SLD and
LHC, one can see that the fermionic three-loop corrections are
negligible compared to the experimental uncertainties. For
example, the direct measurements of the $W$ mass, effective weak mixing angle,
and $Z$ width are $\MW=80.379\pm 0.012\gev$, $\seff{\ell}=0.23152 \pm 0.00016$ and $\Gamma_{\PZ,\rm
tot}=2.4952\pm 0.0023\gev$. These are at least one order of magnitude larger
than the corrections in eqs.~\eqref{dmw}, \eqref{dseff} and \eqref{dgamf}. Our numerical results presented here are different from those in Ref.~\cite{achim} since we use a different renormalization scheme, but within a similar order of magnitude, both of which are expected.
\begin{table}[!ht]
\centering

\begin{tabular}{|l | l | l | l | l | l |}\hline
                       & Current Theory & Main source                                                          & CEPC Exp  & FCC-ee Exp          & ILC Exp  \\
                         \hline
$M_W$[MeV]                            & $4 $              & $\alpha^3$, $\alpha^2 \alpha_s$                                      &$1$        & $1$                     & $2.5-5 $   \\
$\Gamma_Z$[MeV]                        & $0.5 $            &  $\alpha^3$, $\alpha^2\alpha_s$, $\alpha\alpha_s^2$ & $0.5$       & $0.1$ &           $0.8$    \\
$\sin^2{\theta^l_{eff}}$ & $4.3 \times 10^{-5}$   & $\alpha^3$, $\alpha^2 \alpha_s$                                       &       $2.3 \times 10^{-5} $      &          $0.6\times 10^{-5}$               &             $ 10^{-5} $  \\
\hline
\end{tabular}
\mycaption{This table demonstrates the future experimental
accuracies given by CEPC, FCC-ee, and ILC with respect to the three EWPOs, along with the current theoretical
uncertainties due to missing higher order~\cite{cepc,fccee,therr,Irles:2019xny,
zbos }. The methods for estimating the main sources of theory uncertainty are
described in Ref.~\cite{rev1}.}
\label{tab:futcoll}
\end{table}

However, future high-luminosity $e^+e^-$ colliders, such as FCC-ee, CEPC or ILC, are
anticipated to dramatically improve the experimental precision for
these quantities \cite{cepc,fccee,Irles:2019xny}, see Tab.~\ref{tab:futcoll}.
It is evident that the corrections computed in this paper are important for
the physics program of these machines.

Finally, we also wish to study in the impact of the error that was found in the
previous calculation of the fermionic two-loop contribution to the partial decay
widths, $\Delta\overline{\Gamma}_{f,(2)}$, as discussed in section~\ref{calc}. 
Using the inputs from Tab.~\ref{tab:input}, the difference amounts to
\begin{align}
\Delta\overline{\Gamma}_{f,(2)}\Big|_\text{this work} -
\Delta\overline{\Gamma}_{f,(2)}\Big|_\text{Ref.~\cite{mwshort,mwlong}}
&= -N_c^f (v_{f(0)}^2 + a_{f(0)}^2) \,\mz 
\frac{25\alpha^2(3-8\sw^2)^2}{3888\pi\sw^2\cw^2} \\
&= \begin{cases}
-0.0028\mev & \text{for } f = \ell, \\
-0.0056\mev & \text{for } f = \nu, \\
-0.0126\mev & \text{for } f = d, \\
-0.0098\mev & \text{for } f = u, \\
-0.0830\mev & \text{for } f = {\rm tot}.
\end{cases} 
\end{align}
It turns out that the numerical impact is very small, but for the
sake of consistency it is important to identify and correct this error.



\section{Conclusions}
\label{conc}

Table~\ref{tab:futcoll} illustrates the comparison between the current theoretical uncertainties due to missing higher orders and the future experimental targets. For the ILC, the theoretical uncertainty for $M_W$ and the Z-boson width are comparable to the target precision at ILC, but the expected precision for $\seff{\ell}$ at ILC is smaller than the current theoretical uncertainty by roughly a factor of 4. For the other two experiments, the experimental target precision for all observables are mostly smaller than the current theoretical uncertainties (except for the Z-boson partial width at CEPC, which is expected to have an experimental precision comparable to today's theoretical uncertainty).
In particular, the target precision for the effective weak mixing angle at FCC-ee would require an improvement of the current theory error by at least one order of magnitude~\cite{therr}. 

Hence electroweak precision measurements at future $e^+e^-$ colliders, such as CEPC, FCC-ee, and ILC, require the inclusion of three-loop electroweak corrections in
theoretical calculations to match the experimental precision. As a first step,
this article presents results for contributions with three closed fermion loops
for some of the most important electroweak precision observables (EWPOs):
\emph{(i)} The prediction of the $W$ mass from the muon decay rate;
\emph{(ii)} the ratio of vector and axial-vector $Z$-boson couplings; and
\emph{(iii)} $Z$-boson decay rates into different final states.
Corrections with a fixed number of closed fermion loops form a UV-finite and
gauge-invariant subset, and they are enhanced by powers of $\mt$ and the large
multiplicity of light fermion degrees of freedom.

Care must be taken when deriving the counterterms for the $W$ and
$Z$ boson masses. Since these particles have non-negligible decay widths, their
propagator poles are complex. A consistent theoretical definition of the gauge
boson masses is then given by the real part of the complex propagator poles.
Additional complications arise from $\gamma$--$Z$ mixing. 

Given the large size of the final expression, the calculation has been performed
with the help of computer-algebra tools. The numeric size of the leading
fermionic three-loop corrections turns out to be relatively small for all
considered EWPOs, but not negligible for the anticipated precision of CEPC, 
FCC-ee and ILC.

In the course of the calculation, an error was found in literature for the
previously known results for the leading fermionic two-loop corrections to the
$Z$-boson decay width. The numerical impact of this error is found to be very
small.

Experience from electroweak two-loop calculations
\cite{mwshort,mwlong,mw,mwtot,swlept,swlept2,swbb,gz} shows that loop
corrections with maximal number of closed fermion loops and next-to-maximal
number of closed fermion loops can be of comparable size. Thus the availability
of the new results in this paper, while important, does not significantly reduce
the theoretical error estimates given in Refs.~\cite{mwtot,zbos} for the
relevant EWPOs. The theory uncertainties are dominated by other
missing three-loop corrections which will need to be calculated to meet the
goals of future $e^+e^-$ colliders \cite{therr}. At the order of $\alpha^4$, the leading
fermionic four-loop electroweak corrections can be carried out with the same
approach used in this paper, but the expected size is smaller than the target
precision levels in the CEPC, FCC-ee, and ILC/Giga-Z designs.


\section*{Acknowledgments}

This work has been supported in part by the National Science Foundation under
grant no.\ PHY-1820760.


\end{document}